\newcommand{\rem}[1]{}
\documentclass{amsart}
\usepackage{amsfonts,amssymb,amsmath,amsthm,mathrsfs}
\usepackage{url}
\usepackage[dvips]{epsfig}
\newtheorem{thrm}{Theorem}[section]
\newtheorem{lem}[thrm]{Lemma}
\newtheorem{prop}[thrm]{Proposition}

\theoremstyle{definition}
\newtheorem{definition}[thrm]{Definition}

\begin{document}
\author[C.~A.~Mantica and L.~G.~Molinari]
{Carlo~Alberto~Mantica and Luca~Guido~Molinari}
\address{C.~A.~Mantica: 
I.I.S. Lagrange, Via L. Modignani 65, 20161 Milano, 
and I.N.F.N. sez. Milano, Italy -- L.~G.~Molinari: Physics Department `Aldo Pontremoli',
Universit\`a degli Studi di Milano and I.N.F.N. sez. Milano,
Via Celoria 16, 20133 Milano, Italy.}
\email{carlo.mantica@mi.infn.it, luca.molinari@unimi.it}

\begin{abstract}
The simple structure of doubly torqued vectors allows for a natural characterization of doubly twisted  
down to warped spacetimes, as well as Kundt spacetimes down to PP waves. For the first ones the vectors are timelike, 
for the others they are null.
We also discuss some properties, and their connection to hypersurface orthogonal conformal Killing vectors, and null
Killing vectors.
\end{abstract}
\title[Doubly torqued vectors]{Doubly torqued vectors\\ and a classification
of\\ doubly twisted and Kundt spacetimes}
\maketitle
\section{Introduction}
Recently, we introduced timelike doubly torqued vectors \cite{ManMol21}. They provide a simple characterization of
$1+n$ doubly twisted spacetimes, and its subcases of twisted, doubly warped, generalized Robertson-Walker spacetimes.
Remarkably, the same definition of doubly torqued vectors fits in the characterization of Kundt spacetimes: a Kundt 
spacetime is precisely defined by the existence of a null doubly torqued vector, and special cases  
as the Walker and Brinkmann metrics are naturally identified. 
The purpose of this paper is to present such characterizations, that are summarized in the tables of this introduction.

An important variety of spacetimes are foliations with totally umbilical spacelike Riemannian hypersurfaces of dimension $n$, parametrized by time \cite{Ponge}. In proper coordinates, the metric tensor has a $1+n$ block-diagonal structure. Depending on the arguments of the two scale functions $a^2$ and $b^2$, the spacetimes bear different names:
\begin{table}[h!]
\centering
\caption{1+n doubly twisted spacetimes}
\begin{tabular}{ | l |   r  l |} 
\hline
1+n spacetime &  $ds^2 =$ &  \\
\hline
doubly twisted  & $ -b^2(t,{\bf q}) dt^2 $&$ +  a^2 (t,{\bf q})   g_{\mu\nu}^\star ({\bf q}) dq^\mu dq^\nu $   \\
twisted               & $-dt^2 $                 & $     +  a^2 (t,{\bf q})  g_{\mu\nu}^\star ({\bf q}) dq^\mu dq^\nu  $  \\
unnamed1          & $-b^2(t,{\bf q}) dt^2  $&$+ \quad  a^2 (t)  g_{\mu\nu}^\star ({\bf q}) dq^\mu dq^\nu $  \\
unnamed2           & $-b^2(t,{\bf q}) dt^2 $&$ + \qquad\quad   g_{\mu\nu}^\star ({\bf q}) dq^\mu dq^\nu  $ \\
doubly warped   & $-b^2({\bf q}) dt^2  $&$ +  \quad a^2 (t)  g_{\mu\nu}^\star ({\bf q}) dq^\mu dq^\nu $   \\
warped       & $- dt^2         $&$           + \quad  a^2 (t) g_{\mu\nu}^\star ({\bf q}) dq^\mu dq^\nu $     \\
static                  & $-b^2({\bf q}) dt^2     $&$       + \qquad\quad     g_{\mu\nu}^\star ({\bf q}) dq^\mu dq^\nu   $ \\ 
\hline
\end{tabular}
\end{table}

There is a vast literature about them, since the paper by Yano \cite{Yano} in 1940, who introduced doubly
twisted manifolds. Warped $1+n$ spacetimes are also known as generalized Robertson-Walker 
\cite{Alias,BYChen,ManMolGRW}. The table includes spacetimes without name, that naturally emerge in this
classification.\\
The same spacetimes have a tensor characterization, independent of the choice of coordinates, through the 
existence of a timelike-unit vector field $u_i$ that is vorticity-free and shear-free. Besides this description, preferred by physicists, we recently identified another one in terms of a  {\em timelike doubly torqued vector} \cite{ManMol21}:
\begin{align}
\nabla_j \tau_k = \kappa g_{jk} +\alpha_j \tau_k + \tau_j\beta_k \label{doublytorqued}
\end{align}
where $\alpha_k\tau^k=0$ and $\beta_k\tau^k=0$. 
Despite being $u_i=\tau_i/\sqrt {-\tau^2}$, where $\tau^2=\tau_k\tau^k$, the vector $\tau_i$ offers a straightforward classification of the spacetimes (table 2). In some cases, $\alpha_i$ and $\beta_i$ are gradients of scalar functions. In parallel, the vector field $u_i$ gets more and more specialized through requirements on the expansion parameter $\varphi $ and the acceleration $\dot u_i=u^k\nabla_k u_i$.
\begin{table}[h!]
\centering
\caption{Characterizations with timelike doubly torqued and unit vectors}
\begin{tabular}{| l | c  c || c c c |} 
\hline
 $1+n$ spacetime & $\nabla_i\tau_j= $ &  & $\nabla_i u_j  = $ & $\nabla_i\varphi  = $ & \\
\hline
doubly twisted & $\kappa g_{ij}+  \alpha_i\tau_j + \tau_i\beta_j $ & \cite{ManMol21} & $\varphi (u_iu_j+g_{ij})-u_i\dot u_j $ 
& $ -u_i\dot\varphi + v_i $ & \cite{Ferrando} \\
twisted   & $ \kappa g_{ij}+  \alpha_i\tau_j  $ & \cite{Chen2017} & $ \varphi (u_iu_j+g_{ij}) $ & $  -u_i\dot\varphi + v_i $ & \cite{ManMol2017} \\
unnamed1 & $\kappa g_{ij}+  \tau_i\beta_j $ & &  &  & \\
unnamed2 & $ \tau_i\beta_j $ & &  &  & \\
doubly warped  & $\kappa g_{ij}+  \tau_i\partial_j \beta $ &  \cite{ManMol21}  &  & & {}  \\
warped (GRW)  & $\kappa g_{ij} $ &  \cite{Chen2014}  & $\varphi (u_iu_j+g_{ij}) $ & $ -u_i\dot\varphi $  & 
\cite{ManMolGRW} \\
static & $\tau_i\partial_j\beta $  &  &  $-u_i \dot u_j $ & & \\ 
\hline
\end{tabular}
\end{table}

Timelike doubly-torqued vectors extend the characterizations by Bang-Yen Chen of twisted spacetimes in terms of torqued vectors ($\beta_i=0$) and of warped spacetimes in terms of concircular vectors ($\alpha_i=\beta_i =0$). They also identify other spacetimes, that do not have simple description in terms of $u_i$.
The special case $\alpha_i+\beta_i=0$ identifies doubly torqued vectors with hypersurface orthogonal conformal Killing vectors, making contact with literature.

Surprisingly, null doubly torqued vectors exactly match the Newman-Penrose characterization of Kundt spacetimes.
Since $\tau^2=0$ it is $\kappa =0$ in eq.\eqref{doublytorqued}, and a proper rescaling gives a vector $\tau'$: 
\begin{align}
 \nabla_i \tau'_j = \theta \tau'_i\tau'_j +\beta'_i\tau'_j + \tau'_i\beta'_j
 \end{align}
with $\beta'$ the non-null component of $\beta $.  Conditions on $\theta $ and $\beta'$ give special cases, as the Walker anf Brinkmann metric  of PP waves (table 3).
\begin{table}[h!]
\caption{Kundt class spacetimes, and null doubly torqued vectors}
\begin{tabular}{| l |  l  |} 
\hline
Kundt    & $ds^2=H(u,v,{\bf q}) du^2  -2dudv  +2W_\mu (u,v,{\bf q})dudq^\mu   +g_{\mu\nu}(u,{\bf q}) dq^\mu dq^\nu $  \\
  &   $ \nabla_i\tau'_j = \theta\tau'_i\tau'_j+\beta'_i\tau_j + \tau_i\beta'_j  $ \\
\hline 
  &     $ds^2=   H(u,v,{\bf q}) du^2  -2dudv  +2du dq^\mu \partial_\mu [\Phi_0 ({\bf q})+ v\Phi_1({\bf q})]      
+g_{\mu\nu} (u,{\bf q}) dq^\mu dq^\nu$      \\
&        $\nabla_i\tau'_j =\theta\tau'_i\tau'_j+(\partial_i \beta) \tau'_j + \tau'_i (\partial_j\beta)$  \\
\hline
&         $ds^2= H(u,{\bf q}) du^2  -2dudv  +2W_\mu (u,v,{\bf q})dudq^\mu      
+g_{\mu\nu} (u,{\bf q}) dq^\mu dq^\nu $     \\
 &        $\nabla_i\tau'_j =\beta'_i \tau'_j + \tau'_i \beta'_j$   \\
 \hline
Walker      &  $ds^2= H(u,v,{\bf q}) du^2  -2dudv  +2W_\mu (u,{\bf q})dudq^\mu      
+g_{\mu\nu} (u,{\bf q}) dq^\mu dq^\nu     $ \\
 &        $\nabla_i\tau'_j =\theta \tau'_i\tau'_j  $  \\
\hline
Brinkmann       &  $ds^2=  H(u,{\bf q}) du^2  -2dudv  +2W_\mu (u,{\bf q})dudq^\mu      
+g_{\mu\nu} (u,{\bf q}) dq^\mu dq^\nu  $    \\
(PP waves) &          $ \nabla_i\tau'_j =0  $  \\ 
\hline
\end{tabular}
\end{table}

\section{Timelike doubly torqued vectors}
We obtain properties for timelike doubly torqued vectors and revisit the relations among $\tau_i, \kappa , \alpha_i, \beta_i$ and the scale functions $a,b>0$ of the metric, discussed in \cite{ManMol21}, to obtain new results. 
We refer to the coordinate frame where the space components $\tau_\mu $ and $u_\mu$ vanish, as the ``comoving" 
frame.\\
Timelike doubly torqued vectors satisfy the Frobenius condition $\tau_{[i}\nabla_j\tau_{k]}=0$ and are hypersurface orthogonal.

This symmetry is useful:
\begin{prop}
If $\tau_i$ is a timelike doubly torqued vector with  $(\kappa , \alpha_i, \beta_i)$ in eq.\eqref{doublytorqued}, then $\mu \tau_i $ is doubly torqued with $(\mu \kappa, \alpha_i+\partial_i\mu/\mu, \beta_i)$ provided that $\tau^k\partial_k \mu =0$.\\ In the
comoving frame ($\tau_\mu=0$) the condition means that $\partial_t\mu =0$.\\
If $\alpha_i=\partial_i \alpha $ (orthogonal to $\tau_i$), then a rescaling of $\tau_i$ brings it to $\alpha_i=0$.
\end{prop}
\noindent
Let us enquire when $\alpha_i$ is a gradient, i.e. is closed. Contraction of \eqref{doublytorqued} with $\tau^k$ gives:
\begin{align}
\alpha_j = \nabla_j \log\sqrt { -\tau^2} - \kappa \frac{\tau_j}{\tau^2}   \label{eqalpha}
\end{align}
The evaluation of $\nabla_i\alpha_j$ gives the useful identity
\begin{align}
(\nabla_i \alpha_j -\nabla_j\alpha_i) \tau^2 =\tau_i(\nabla_j \kappa -\kappa \alpha_j -\kappa\beta_j) - \tau_j (\nabla_i \kappa -\kappa\alpha_i -\kappa \beta_i)   \label{NEWEQ}
\end{align}
\begin{prop}
$\alpha_j$ is closed if and only if $\nabla_j \kappa -\kappa \alpha_j -\kappa\beta_j $ is parallel to $\tau_j$.
\end{prop}

In the comoving frame $\tau_\mu=0$, $\alpha_0 =\beta_0=0$, with the Christoffel symbols listed in appendix,
eq.\eqref{doublytorqued} for doubly torqued vectors becomes ($\mu=1,...,n$):
\begin{align*}
\begin{array}{cc}
 \partial_t \tau_0 -\tau_0\partial_t \log b = -\kappa b^2, \\
\partial_\mu \tau_0- \tau_0\partial_\mu \log b =\tau_0\alpha_\mu \\
-\partial_\mu \log b =\beta_\mu, \\
- \tau_0\partial_t\log a = \kappa b^2
\end{array}
\end{align*}
%
The following propositions concern the two unnamed spacetimes, respectively, and their subcases:
\begin{prop}
In a doubly twisted spacetime, if $\alpha_i=0$ (or $\alpha_i$ is a gradient orthogonal to $\tau $) 
then $a^2(t)$ only depends on time.
\begin{proof}
If $\alpha_\mu=0$ the second equation gives $\tau_0 (t,{\bf q}) = F(t) b(t,{\bf q})$ with some function 
$F$.  The first and last equations give $\partial_t \log a = (\partial_t F)/F(t)$.
\end{proof}
\end{prop}

\begin{prop}\label{prop25}
In a doubly twisted spacetime, $\kappa =0$ if and only if $a^2$ only depends on ${\bf q}$ (and may be included in $g^\star_{\mu\nu} ({\bf q})$).\\ 
Then $\alpha_i$ is a gradient (and can be absorbed to zero) and $\tau^2$ is independent of
time. 
\begin{proof}
The last equation gives $a^2$ that only depends on ${\bf q}$ if and only if $\kappa =0$. 
The first one gives $\tau_0=C({\bf q})b(t,{\bf q})$,
and the second one results in $\alpha_\mu =\partial_\mu \log C({\bf q})$. Then $\alpha_i$ is a spacetime gradient. Eq.
\eqref{eqalpha} gives $\alpha_i =\nabla_i \log \sqrt{-\tau^2}$. In the comoving frame $\alpha_0 =0$ so that
$\tau^2 $ is independent of time.
\end{proof}
\end{prop}




\section{Timelike hypersurface orthogonal conformal Killing vectors} 
We show that timelike doubly-torqued vectors with $\alpha_i+\beta_i=0$ coincide with hypersurface orthogonal conformal Killing vectors (\cite{Hall} Ch.11, \cite{Stephani} pp.69, 564). We revisit in this
light some theorems, and give new ones.
\begin{definition}
$\xi_i$ is a conformal Killing vector if $\nabla_i\xi_j + \nabla_j \xi_i = 2\kappa g_{ij}$ or, equivalently,
$\nabla_i \xi_j = \kappa g_{ij} + F_{ij}$ with $F_{ij}=-F_{ji}$.  It is a Killing vector if also $\kappa=0 $. 
\end{definition}
\begin{lem}
A timelike conformal Killing vector $\xi_i$ is hypersurface orthogonal if and only if:
$ F_{jk} = \alpha_j \xi_k - \xi_j \alpha_k$, $\alpha_k\xi^k=0 $, i.e.
\begin{align}
 \nabla_i \xi_j =\kappa g_{ij} + \alpha_j \xi_k - \xi_j \alpha_k  \label{CKT}
 \end{align}
\begin{proof}
By the Frobenius theorem, a vector is hypersurface orthogonal if and only if $0=\xi_{[i}\nabla_j \xi_{k]} = 
\xi_i (\nabla_j \xi_k -\nabla_k \xi_j)+$cyclic permutations i.e.
$ \xi_i F_{jk} + \xi_j F_{ki} + \xi_k F_{ij} =0$.
A contraction with $\xi^i$ gives $\xi^2 F_{jk}+\xi_j (F_{ki}\xi^i) - \xi_k (F_{ji}\xi^i)=0$. It is always possible to choose $\alpha_k\xi^k =0$, as $\alpha_k - \alpha_j\xi_k\xi^j/\xi^2$ does the job.
\end{proof}
\end{lem}

\begin{prop}
Doubly torqued vectors with $\alpha_i =-\beta_i$ are hypersurface orthogonal conformal Killing vectors. 
They are hypersurface orthogonal Killing vectors if also $\kappa =0$.
\end{prop}

In ref.\cite{ManMol21} we showed that a doubly twisted spacetime is doubly warped if and only if $\alpha_i=\partial_i
\alpha  $ and $\beta_i =\partial_i\beta $ in \eqref{doublytorqued} (see table 2). Since they are both orthogonal to 
$\tau $ we may rescale $\tau $ such that $\alpha_i=-\partial_i \beta$ and obtain $\nabla_i\tau_j =\kappa g_{ij} -(\partial_i \beta)\tau_j + \tau_i(\partial_j\beta)$, a conformal Killing vector. Therefore:
\begin{prop} 
A spacetime is doubly warped if and only if it is equipped with a hypersurface orthogonal conformal Killing vector 
with closed vector $\alpha_i$.
\end{prop}

With $\alpha_i =-\beta_i$ in \eqref{NEWEQ}, we read that $\alpha_j $ is closed if and only if $\nabla_j \kappa $ is proportional to $\tau_k$. Therefore, we have the statement (Theorem 1 in \cite{Ramos}): 
 {\em A spacetime is doubly warped if and only if it is equipped with a hypersurface orthogonal conformal Killing vector with $\partial_i\kappa $ parallel to $\xi_i$}. \\
 Moreover, if $\tau $ is closed ($\alpha_i =\beta_i$) then $\alpha_i=\beta_i=0$: the spacetime is 
generalized Robertson-Walker (Cor. 2 in \cite{Ramos}).

A doubly torqued vector with $\kappa =0$, $\alpha_i=-\beta_i$ is a hypersurface orthogonal Killing vector. Since
$\alpha_i $ and $-\alpha_i$ are gradients (Prop.\ref{prop25}), the spacetime is doubly warped. Then $a^2$ is a function of $t$ and $b^2$ is a function of ${\bf q}$. $\kappa =0$ means that $\partial_t a=0$ i.e. $a$ is a constant. The metric 
$ ds^2 = -b^2({\bf q})dt^2 + a^2 g_{\mu\nu}^\star ({\bf q}) dq^\mu dq^\nu $
 has the form of a static spacetime \cite{Stephani} p.283.

\section{Null doubly torqued vectors and Kundt spacetimes}
A Kundt spacetime is defined by the presence of a null geodesic congruence that is expansion-free,
shear-free, and twist-free \cite{Stephani} Ch.31, \cite{Kundt,Coley09,Podolsky09,Ortaggio13}. We show that it precisely means that it admits a doubly torqued null vector field.\\
We begin with some facts on null doubly torqued vectors.

The contraction of $\nabla_i\tau_j =\kappa g_{ij} + \alpha_i\tau_j + \tau_i\beta_j$ with $\tau^j$ gives $\kappa =0$. Then:
\begin{align}
\nabla_i \tau_j = \alpha_i \tau_j + \tau_i\beta_j, \quad \alpha_k\tau^k=0, \; \beta_k\tau^k=0. \label{NDT}
\end{align}
Contraction with $\tau^j $ gives that $\tau $ is geodesic: $\tau^i\nabla_i \tau_j =0$.\\ 
For null vectors one considers the optical scalars \cite{Poisson}:
\begin{align}
\Theta= \frac{1}{d-2} \nabla_k \tau^k , \quad \omega^2 = -\nabla_{[k}\tau_{j]} \nabla^k\tau^j , \quad \sigma^2 = \nabla_{(k}\tau_{j)}\nabla^k \tau^j - (d-2)\Theta^2
\end{align}
where $d$ is the dimension of spacetime.
It is simple to prove that all the three optical scalars vanish for null doubly torqued vectors. In particular, the
vanishing of the twist ($\omega^2=0$) is the condition for $\tau $ to be hypersurface orthogonal.

Since  $\tau^2=0$, $\alpha_i= a\tau_i +\alpha'_i$ where $\alpha'$ is a spacelike vector orthogonal to $\tau $, 
and $\beta =b\tau_i+\beta_i'$. Then, for a null doubly torqued vector, with $\theta=a+b$, it is
\begin{align}
\nabla_i\tau_j = \theta \tau_i \tau_j + \alpha'_i\tau_j + \tau_i \beta'_j  \label{DTK}
\end{align}

We now turn to Kundt spacetimes and show that \eqref{DTK} is precisely the equation for the congruence. Let $\ell_i$ be the geodesic null congruence, and $n_i$ a second null vector field with $n_i \ell^i=-1$. $\hat h_{ij} = g_{ij} + \ell_i n_j + n_i\ell_j$ is the projection on the space orthogonal to $\ell $ and $n$.
Consider the decomposition
\begin{align*}
\nabla_i \ell_j =& (\hat h_i^l - \ell_in^l - n_i \ell^l)(\hat h_j^m -\ell_jn^m - n_j \ell^m) \nabla_l \ell_m \\
=&(\hat h_i^l - \ell_in^l)(\hat h_j^m -\ell_jn^m) \nabla_l \ell_m \\
=&\hat h_i^l \hat h_j^m \nabla_l \ell_m + \ell_i\ell_j (n^ln^m \nabla_l \ell_m) - \hat h_i^l \ell_k n^m \nabla_l \ell_m
- \ell_in^l \hat h_j^m \nabla_l \ell_m 
\end{align*}
The omitted terms contain $\ell^l\nabla_l \ell_m=0$ (the field is geodesic) and $\ell^m\nabla_l \ell_m=0$.
The first term is the projection onto the subspace of dimension $d-2$ orthogonal to $\ell_i$ and $n_i$, and is decomposed into expansion, shear and twist: 
\begin{align*}
\hat h_i^l \hat h_j^m \nabla_l \ell_m = \frac{\nabla_l \ell^l}{d-2} \hat h_{ij} +\hat\sigma_{ij} +\hat\omega_{ij}
\end{align*}
For Kundt spacetimes these terms are zero, and we have the known statement (we shift to the letter $\tau_i $):
\begin{gather*}
\nabla_i \tau_j  = (n^ln^m \nabla_l \tau_m) \tau_i\tau_j  - (\hat h_i^l n^m \nabla_l \tau_m)\tau_j  -\tau_i (\hat h_i^m n^l \nabla_l \tau_m)
\end{gather*}
\begin{thrm}
A spacetime is Kundt if and only if there is a doubly torqued null vector field, eq.\eqref{NDT} or \eqref{DTK}.
\end{thrm}
The property $\lambda \tau_i =\nabla_i f$ (hypersurface orthogonality) offers a rescaling of $\tau $ that makes it 
a closed vector:
\begin{prop}
The vector $\tau'_i =\lambda \tau_i $ is null doubly torqued, closed, and 
\begin{align}
\nabla_i\tau'_i  = \theta \tau'_i \tau'_j + \beta'_i \tau'_j + \tau'_i \beta'_j  \label{DTK2}
\end{align}
where the vector $\beta'$ is the component of $\beta $ not aligned with $\tau $.
\begin{proof}
The evaluation gives: $\nabla_i\tau'_j = (\alpha_i +\partial_i \lambda/\lambda)\tau'_j + \tau'_i \beta_j$.
Since $\tau'_i $ is closed, it is  $(\alpha_i -\beta_i +\partial_i \lambda/\lambda )\tau'_j = 
(\alpha_j -\beta_j +\partial_j \lambda/\lambda )\tau'_i$. Then: $\alpha_i +\partial_i \lambda/\lambda =\beta_i+\gamma \tau'_i$ and  $\nabla_i\tau'_j = \gamma \tau'_i \tau'_j + \beta_i\tau'_j + \tau'_i \beta_j$. Next, being 
$\beta_i\tau^i=0$ and $\tau $ null, it is $\beta =b\tau_i +\beta'_i$. The expression is obtained.
\end{proof}
\end{prop}
The metric of a Kundt spacetime in coordinates adapted to the null vectors is:
\begin{align}
ds^2 = H(u,v,{\bf q}) du^2 -2dudv + 2W_\mu(u,v,{\bf q}) du dq^\mu +g_{\mu\nu} (u,{\bf q})dq^\mu dq^\nu 
\end{align}
The coordinates $u$ and $v$ refer to the subspace spanned by $\tau_i$ and $n_i$, where $\tau_u=-1$, $\tau_v=0$, $\tau_\mu =0$, $\alpha'_u=\beta'_u=0$. Eq.\eqref{DTK} gives the following relations:
\begin{align}
\theta = \frac{1}{2}\frac{\partial H}{\partial v}, \quad \alpha'_v=\beta'_v =0, \quad \alpha'_\mu=\beta'_\mu =
-\frac{1}{2}\frac{\partial W_\mu}{\partial v}
\end{align}
It turns out that the metric is evaluated with the vector \eqref{DTK2}.\\
We have three special cases:
\begin{enumerate}
\item $\partial H/\partial v=0$ corresponds to $\theta =0$
\item $\partial W_\mu/\partial v=0$, i.e. $\alpha'_i=\beta'_i=0$. It is $\nabla_i \tau_j = \theta \tau_i\tau_j$. 
This recurrent case gives the Walker metric  \cite{Leistner}.
\item $\partial H/\partial v=0$ and $\partial W_\mu/\partial v=0$ equivalent to $\theta=0$, $\alpha'_i=\beta'_i=0$. This case 
gives the Brinkmann metric (PP wave, i.e. plane-fronted waves with parallel propagation) \cite{Coley03,Ortaggio13}.
\end{enumerate} 

Another special case is $\beta' $ closed. The equation $\nabla_i\beta'_j = \nabla_j \beta'_i $ gives: 
1) $\partial_\mu \beta'_\nu=\partial_\nu\beta'_\mu $ i.e. $W_\mu = \partial_\mu \Phi (u,v,{\bf q})$ for some potential; 
2) $\partial_u \beta'_\mu =0$, then $\Phi $ does not depend on $u$;
3) $\partial_v \beta'_\mu = 0 $,   then $\Phi $ is a linear function of $v$. 
In summary: $\beta' $ closed implies $W_\mu (v,{\bf q}) = \partial_\mu \Phi_0 ({\bf q}) + v \partial_\mu\Phi_1 ({\bf q})$, 
(in table 3).\\
This case is realized in the solutions of the Einstein-Maxwell equations in vacuo, or with electromagnetic field aligned to
$\tau $ ($F_{ij}\tau^j \propto \tau_i$), or with the cosmological constant. For this problem
$H$ is a quadratic function of $v$ (eq.77 and 112 in  \cite{Podolsky09}). 

\section{Null hypersurface orthogonal Killing vectors}
In analogy with timelike vectors, we consider null doubly torqued vectors with $\alpha_i=-\beta_i$. They coincide with (hypersurface orthogonal) null Killing vectors, and describe a subclass of Kundt spacetimes \cite{Dautcourt}. 
\begin{prop} A null hypersurface orthogonal Killing vector is a doubly torqued vector with $\alpha_i =-\beta_i$.\\
A null doubly torqued vector $\nabla_i\tau_j=\alpha_i\tau_j -\tau_i\alpha_j$ is a Killing vector.
\begin{proof}
The hypothesis are: $\nabla_i \tau_j =F_{ij}$ ($F_{ij}=-F_{ji}$) and $\tau_i=\lambda \nabla_i f$. Then:
$F_{ij}= (\nabla_i\lambda )\nabla_j f + \lambda \nabla_i\nabla_j f$. Subtraction of $F_{ji}$ gives
$F_{ij} = \frac{1}{2}\frac{\nabla_i\lambda}{\lambda } \tau_j -  \frac{1}{2}\frac{\nabla_j\lambda}{\lambda } \tau_i $. Since 
$F_{ij}\tau^j=0$, the vector $\tau $ is doubly torqued with $\alpha_i=-\beta_i$.\\
A doubly torqued vector is hypersurface orthogonal and, if $\beta_i=-\alpha_i$ it is $\nabla_i\tau_j + \nabla_j \tau_i=0$ i.e.
$\nabla_i\tau_j = F_{ij}=-F_{ji}$.
\end{proof}
\end{prop}
The metric in $d=4$ is given in \cite{Stephani} p.380.
If $\tau_i$ is also closed, then $\nabla_i\tau_j=0$ and PP waves are obtained.

\section{Curvature tensors}
The integrability conditions for a null or timelike doubly torqued vector are:
\begin{align}
R_{jklm}\tau^m 
=g_{kl}(\nabla_j \kappa -\kappa \alpha_j) - g_{jl}(\nabla_k \kappa -\kappa\alpha_k) +(\nabla_j \alpha_k -\nabla_k\alpha_j) \tau_l   \label{RiemTau}\\
 +\tau_k(\nabla_j\beta_l -\beta_j\beta_l)- \tau_j(\nabla_k\beta_l -\beta_k\beta_l) \nonumber
\end{align}
The contraction of the Ricci tensor with $\tau^m $ is obtained:
\begin{align}
R_{km}\tau^m 
= -(n-1)\nabla_k \kappa  + \kappa(n\alpha_k +\beta_k) +\tau^j\nabla_j \alpha_k +\tau_k(\alpha^j\beta_j +\nabla_j\beta^j) 
\label{RicciC}
\end{align}
Then, a null $\tau $ is eigenvector if and only if $\tau^j\nabla_j \alpha_k \propto \tau_k$.
\begin{lem}
For null doubly torqued vectors:
\begin{align*}
&\tau_i \nabla_j (\alpha_k-\beta_k) + \tau_j \nabla_k (\alpha_i-\beta_i) + \tau_k \nabla_i (\alpha_j -\beta_j) =0\\
&\tau^k \nabla_k (\alpha_i -\beta_i) = \tau_i (\alpha^k\beta_k -\beta^2) 
\end{align*}
\begin{proof} The first Bianchi identity $R_{jklm}+R_{kljm}+R_{ljkm}=0$  is contracted with $\tau^m$ and the
expressions \eqref{RiemTau} are inserted, with $\kappa =0$.\\ 
Contraction with $\tau^k$ gives the other identity.
\end{proof}
\end{lem}
The property of Weyl or Riemann compatibility for vectors and symmetric tensors is presented in \cite{ManMolGray}. 
Riemann compatibility implies Weyl compatibility. 
\begin{thrm}
A timelike doubly torqued vector is Weyl compatible:
\begin{align}
\tau_i C_{jklm}\tau^m + \tau_j C_{kilm}\tau^m + \tau_k C_{ijlm}\tau^m =0 \label{WeylC}
\end{align}
A null doubly torqued vector with $\alpha_i $ closed or with $\beta_i = C\alpha_i$ with $C\neq 1$ a constant, 
is Riemann compatible 
\begin{align}
\tau_i R_{jklm}\tau^m + \tau_j R_{kilm}\tau^m + \tau_k R_{ijlm}\tau^m =0 \label{RiemC}
\end{align}
and is an eigenvector of the Ricci tensor.
\begin{proof}
Multiplication of \eqref{RiemTau} by $\tau_i$ and a cyclic sum give:
\begin{align*}
&\tau_i R_{jklm}\tau^m + \tau_j R_{kilm}\tau^m + \tau_k R_{ijlm}\tau^m \\
&= [\tau_i (\nabla_j \alpha_k-\nabla_k\alpha_j) +\tau_j (\nabla_k \alpha_i-\nabla_i\alpha_k) +\tau_k (\nabla_i \alpha_j-\nabla_j\alpha_i)]\tau_l \\
&\quad - g_{il}[\kappa (\tau_j \alpha_k - \tau_k\alpha_j) - (\tau_j \nabla_k \kappa - \tau_k \nabla_k \kappa )] +\\
&\quad - g_{jl}[\kappa (\tau_k \alpha_i - \tau_i\alpha_k) - (\tau_k \nabla_i \kappa - \tau_i \nabla_k \kappa )] +\\
&\quad - g_{kl}[\kappa (\tau_i \alpha_j - \tau_j\alpha_i) - (\tau_i \nabla_j \kappa - \tau_j \nabla_i \kappa ]
\end{align*}
If $\tau_i$ is null it is $\kappa =0$. If also $\nabla_j \alpha_k=\nabla_k\alpha_j$ or if $\beta_i=C\alpha_i$ then the cyclic sum is zero (in the second case, use the Lemma).\\
The contraction of \eqref{RiemC} with $g^{jl}$ gives $\tau_i R_{km}\tau^m =  \tau_k R_{im}\tau^m $. Then $\tau $ is an eigenvector
of the Ricci tensor.\\
Let $\tau_i$ by timelike. The contraction of the Weyl tensor with $\tau $ is:
\begin{align*}
C_{jklm}\tau^m =& R_{jklm}\tau^m + \frac{1}{n-2} [\tau_j R_{kl}-\tau_k R_{jl}] \\
&+ \frac{1}{n-2}g_{kl}\left[R_{jm}\tau^m - \frac{R\tau_j}{n-1}\right ] -  \frac{1}{n-2}g_{jl}\left[R_{km}\tau^m - \frac{R\tau_k}{n-1}\right ] 
\end{align*}
Multiplication by $\tau_i$ and a ciclic sum give:
\begin{align*}
&\tau_i C_{jklm}\tau^m + \tau_j C_{kilm}\tau^m + \tau_k C_{ijlm}\tau^m \\
&= [\tau_i (\nabla_j \alpha_k-\nabla_k\alpha_j) +\tau_j (\nabla_k \alpha_i-\nabla_i\alpha_k) +\tau_k (\nabla_i \alpha_j-\nabla_j\alpha_i)]\tau_l \\
&+\tfrac{1}{n-2} g_{kl}\{(\tau_i R_{jm}-\tau_j R_{im})\tau^m  - (n-2) [\kappa (\tau_i \alpha_j - \tau_j\alpha_i) - (\tau_i \nabla_j \kappa - \tau_j \nabla_i \kappa ]\} \\
&+\tfrac{1}{n-2} g_{jl}\{(\tau_k R_{im}-\tau_i R_{km})\tau^m -(n-2)[\kappa (\tau_k \alpha_i - \tau_i\alpha_k) - (\tau_k \nabla_i \kappa - \tau_i \nabla_k \kappa )]\}\\
&+\tfrac{1}{n-2} g_{il}\{(\tau_j R_{km}-\tau_k R_{jm})\tau^m -(n-2) [\kappa (\tau_j \alpha_k - \tau_k\alpha_j) - (\tau_j \nabla_k \kappa - \tau_k \nabla_k \kappa )]\}
\end{align*}
The contraction of the Ricci tensor with $\tau $ is \eqref{RicciC}.
The cyclic sum for the Weyl tensor simplifies:
\begin{align*}
&\tau_i C_{jklm}\tau^m + \tau_j C_{kilm}\tau^m + \tau_k C_{ijlm}\tau^m \\
&= [\tau_i (\nabla_j \alpha_k-\nabla_k\alpha_j) +\tau_j (\nabla_k \alpha_i-\nabla_i\alpha_k) +\tau_k (\nabla_i \alpha_j-\nabla_j\alpha_i)]\tau_l \\
&+\tfrac{1}{n-2} g_{kl}[\tau_i (-\nabla_j\kappa +\kappa (2\alpha_j+\beta_j)+\tau^m\nabla_m\alpha_j)
-\tau_j (-\nabla_i\kappa +\kappa (2\alpha_i+\beta_i)+\tau^m\nabla_m\alpha_i)]\\
&+\tfrac{1}{n-2} g_{jl} [\tau_k (-\nabla_i\kappa +\kappa (2\alpha_i+\beta_i)+\tau^m\nabla_m\alpha_i)
-\tau_i (-\nabla_k\kappa +\kappa (2\alpha_k+\beta_k)+\tau^m\nabla_m\alpha_k)]   \\
&+\tfrac{1}{n-2} g_{il} [\tau_j (-\nabla_k\kappa +\kappa (2\alpha_k+\beta_k)+\tau^m\nabla_m\alpha_k)
-\tau_k (-\nabla_j\kappa +\kappa (2\alpha_j+\beta_j)+\tau^m\nabla_m\alpha_j)]
\end{align*}
For timelike vectors, contraction of \eqref{RiemTau} by $\tau^l\tau^k$ gives:
\begin{align*}
0=&\tau^2 (\nabla_j \kappa -\kappa \alpha_j) - \tau_j\tau^k \nabla_k \kappa  +\tau^2 \tau^k(\nabla_j \alpha_k -\nabla_k\alpha_j)  +\tau^2 \tau^l \nabla_j\beta_l - \tau_j\tau^l\tau^k\nabla_k\beta_l \\
=&\tau^2 [\nabla_j \kappa -\kappa (2\alpha_j+\beta_j) - \tau^k\nabla_k\alpha_j] - \tau_j(\tau^k \nabla_k \kappa +\tau^2  \alpha^k\beta_k)
\end{align*}
With this identity and \eqref{NEWEQ} the cyclic sum is zero.
\end{proof}
\end{thrm}
Some remarks:\\
- For a timelike doubly torqued vector: $C_{jklm}\alpha^j \beta^k\tau^m =0$. \\
- Weyl compatibility \eqref{WeylC} guarantees that all doubly twisted spacetimes are purely electric \cite{Ortaggio}.\\
- Null hypersurface orthogonal Killing vectors are Riemann compatible. \\
- A Kundt spacetime with Weyl compatible vector $\tau $ is type II(d) in the Bel-Debever classification (table 4 in \cite{Ortaggio13}).

\section{Conclusions}
We showed that the structure of doubly torqued vector is the necessary and sufficient condition for the spacetime
to be doubly twisted (timelike vector) or a Kundt spacetime (null vector). A simple classification of relevant
subcases follows, with connection to other characterizations in terms of Killing or conformal Killing vectors.

\section*{Appendix}
\noindent
The Christoffel symbols for the doubly-twisted metric:
\begin{gather*}
\Gamma_{0,0}^0=\frac{\partial_t b}{b}, \quad 
\Gamma_{\mu,0}^0=\frac{b_\mu }{b},\quad  \Gamma_{0,0}^\mu= \frac{bb^\mu}{a^2},\quad \Gamma^\rho_{\mu,0} = 
\frac{\partial_t a}{a} \delta^\rho_\mu, \quad 
\Gamma^0_{\mu,\nu} = \frac{a\partial_t a}{b^2} g^*_{\mu\nu}, \\
\Gamma^\rho_{\mu,\nu} = \Gamma^{*\rho}_{\mu,\nu} + \frac{a_\nu}{a}  \delta^\rho_\mu + \frac{a_\mu}{a}  \delta^\rho_\nu - \frac{a^\rho}{a}   g^*_{\mu\nu} 
\end{gather*}
where $a_\mu = \partial_\mu a$ and $a^\mu = g^{*\mu\nu} a_\nu $, and the same is for $b$.\\
The Christoffel symbols for the Kundt metric (that are needed in this paper. Taken from \cite{Podolsky09}) :
\begin{gather*}
\Gamma_{u,u}^u=\frac{1}{2}\frac{\partial H}{\partial v}, \quad \Gamma^u_{\mu,u} = \frac{1}{2} \frac{\partial W_\mu}{\partial v}\quad 
\Gamma_{u,v}^u=\Gamma_{v,v}^u=\Gamma_{\mu ,v}^u=\Gamma^u_{\mu,\nu} = 0
\end{gather*}

\section{Declarations}
Not applicable

\end{document}